\documentclass[twoside,final]{IEEEtran}
\usepackage{amsmath}
\usepackage{amssymb}
\usepackage[dvips]{graphicx}
\usepackage{cite}
\usepackage{amsthm}

\title{Asymptotic Error Performance Analysis of Spatial Modulation under Generalized Fading}

\author{\vspace{0.5cm}Kostas~P.~Peppas \IEEEmembership{Senior Member,~IEEE}, Martin Zamkotsian, Fotis Lazarakis and Panayotis G. Cottis
\thanks{K. P. Peppas, M. Zamkotsian and F. Lazarakis are with the  Laboratory of Wireless Communications, Institute of Informatics and Telecommunications, National Centre for Scientific Research--``Demokritos," Patriarhou Grigoriou and Neapoleos, Agia Paraskevi, 15310, Athens, Greece (e-mail: {kpeppas, mzamko, flaz}@iit.demokritos.gr).
M. Zamkotsian and P. G. Cottis are with the National Technical University of Athens, School of Electrical and Computer Engineering, 15773, Athens, Greece
(e-mail: pcottis@central.ntua.gr)
}
}
\begin{document}
\maketitle
\vspace{0.5cm}
\begin{abstract}
This letter presents a comprehensive framework analyzing the asymptotic error performance
of a multiple-input-multiple-output (MIMO) wireless system employing spatial modulation (SM)
with maximum likelihood detection and perfect channel
state information.
Generic analytical expressions for the diversity and coding gains are deduced that reveal fundamental properties of MIMO SM systems.
The presented analysis can be used
to obtain closed-form upper bounds
for the average bit error probability (ABEP) of MIMO SM systems under generalized fading which become asymptotically tight in the high signal-to-noise ratio (SNR) region.
\end{abstract}

\section{Introduction}\label{Sec:Intro}
\newtheorem{proposition}{Proposition}
Spatial modulation (SM) is an efficient, low-complexity transmission technique for multiple-input-multiple-output (MIMO) wireless systems
which achieves a spatial multiplexing gain, at the same time avoiding inter-channel interference without
requiring synchronization between the transmit antennas
\cite{J:RenzoProceedingsIEEE,J:SMRenzoSurvey}. A fundamental concept in SM is
the three-dimensional constellation diagram \cite{J:SMRenzoSurvey} where
each spatial constellation point, corresponding to the transmit antenna index, defines an independent complex plane of
signal constellation points.
When the information carrying entity is solely the transmit-antenna index, SM is reduced
to the space shift keying (SSK) modulation, where
a single transmit-antenna is activated each time to transmit a symbol.
%The activated antenna index is then employed to implicitly convey information.
%the incoming bitstream is encoded into the index of the unique antenna each time switched
%on for transmission while the others are switched off.

Several analytical frameworks assessing the error performance of SM systems over fading channels are available in the technical literature. %Interested readers are referred to \cite{J:RenzoRice, J:RenzoNakagami, J:RenzoPCI, J:SMRenzoHaasTVT, J:Peppas_13} and references therein.
For example, \cite{J:RenzoNakagami} and \cite{J:RenzoRice} employ a moment generating function (MGF) based approach to evaluate the average bit error probability (ABEP) of
SSK in the presence of Nakagami-$m$ and Rice fading. The MGF-based approach presented in these works is extended to the most general case of SM in \cite{J:SMRenzoHaasTVT}, where tight error performance bounds are deduced.
In a recent work \cite{J:Peppas_13}, a generic approach for the performance of SSK was proposed, assuming generalized fading envelopes and uniformly distributed channel phases.

The above cited frameworks provide an exact performance analysis of SM systems over the entire signal-to-noise ratio (SNR) region; however, single integrals
with finite or infinite limits have to be readily evaluated via numerical integration to this end.
Moreover, these frameworks do not provide enough insight into the parameters affecting system performance in terms of diversity and coding gains. In an attempt to bridge this gap, closed-form expressions for the asymptotic performance of SM systems are provided in \cite{J:SMRenzoHaasTVT} and \cite{J:RenzoAsym}.
%In that work, closed-form expressions for the diversity and coding gains of MIMO SSK and SM systems were derived.

Motivated by the above cited works, the objective of the current letter is twofold: a) To deduce
closed-form upper bounds for the ABEP of SM MIMO systems operating over generalized fading environments which become asymptotically tight in the high SNR region,
and b) to provide important considerations about the diversity and coding gains of SM in the presence of generalized fading.
The proposed analysis is tested and verified by numerically evaluated results accompanied
with Monte Carlo simulations as well as by reducing
them to several special cases available in the literature.
\parskip=0pt % disables indentation
\section{Mathematical Tools} \label{Sec:Tools}
%According to \cite{J:RenzoNakagami}, the
%ABEP of a $2 \times 1$ MISO system with Maximum-Likelihood Detection and full Channel
%State Information at the receiver can be evaluated as
%\begin{equation}\label{Eq:ABEP2x1}
%\overline{P} = \int_0^{\infty}Q(|r|\sqrt{\overline{\gamma}})f_{|r|}(r)dr
%\end{equation}
%where $r = a_2\exp(\jmath \phi_2)- a_1\exp(\jmath \phi_1)$ and
%$a_i$ and $\phi_i$ are the channel envelopes and phases of the $i$-th link, $i = 1,2$.
%For a $2 \times N_r$ MIMO system,
%the ABEP can be obtained from \cite[Eq. (11)]{J:RenzoRice} as
In this section, new mathematical tools are presented that
simplify the performance evaluation of SM systems.
According to \cite{J:SMRenzoHaasTVT}, the evaluation of the ABEP of SM requires the solution of integrals of the form
%\footnote{
%In this paper, the Probability Density Function (PDF) of a random variable (RV) $X$ is denoted as $f_X(\cdot)$ and its corresponding Moments Generating Function (MGF) as $\mathcal{M}_X(\cdot)$. Moreover,
%$\jmath = \sqrt{-1}$, $|\cdot|$ is the magnitude of a complex number, $\Gamma(\cdot)$ is the gamma function \cite[Eq. (8.310/1)]{B:Gradshteyn_00}, $\mathbb{E}\langle \cdot \rangle$ denotes expectation,
%$\mathcal{H}_{k,R}\{f(r)\}$ is the $k$-th order Hankel transform of function $f(r)$,
%$H \,\substack{ m , n \\ p , q}[\cdot]$ is the Fox's H-function \cite[Eq. (8.3.1)]{B:Prudnikov3}, $G \,\substack{ m , n \\ p , q}[\cdot]$ is the Meijer's G-function \cite[Eq. (9.301)]{B:Gradshteyn_00}
%and $\null_pF_q(\cdot)$ is the generalized hypergeometric function
%\cite[Eq. (9.14.1)]{B:Gradshteyn_00}. .
%}
\begin{align}\label{Eq:EqIint}
\mathcal{I}(A, L) = \frac{1}{\pi}\int_0^{\pi/2}\prod_{\ell=1}^{L}\left[ \mathcal{M}_{Z_{\ell}}\left(\frac{A}{2\sin^2\theta}\right)\right]\mathrm{d}\theta, \,A>0
%\end{equation}
\end{align}
%of the MGF of the random variables\footnote{Note that symbols with ˜$\tilde{.}$ refer to the actual message that is
%transmitted, while symbols without $\tilde{.}$ denote the trial message that is tested by
%the ML detector, in a similar fashion as in \cite{J:SMRenzoHaasTVT}.}
%$\gamma_{(n_t,\tilde{n}_t)} = \sum_{n=1}^{N_t} |\alpha_{\tilde{n}_t, n_r}- \alpha_{{n}_t, n_r}|^2$ and $\gamma_{(n_t,\chi_\ell, \tilde{n}_t, \chi_{\tilde{\ell}})} = \sum_{n=1}^{N_t} |\chi_{\tilde{\ell}}\alpha_{\tilde{n}_t, n_r}- \chi_\ell \alpha_{{n}_t, n_r}|^2$, for $\text{ABEP}_\text{spatial}$ and $\text{ABEP}_\text{joint}$, respectively \cite[Eqs. (7), (8)]{J:SMRenzoHaasTVT} where $\alpha_{{n}_t, n_r}$ denotes the complex channel gain from the $n_t$th transmit to the $n_r$th receive antenna.
%It becomes evident that the evaluation of the $\text{ABEP}_\text{spatial}$ and $\text{ABEP}_\text{joint}$ requires the evaluation of integrals of the form
where $\mathcal{M}_{Z_{\ell}}(\cdot)$ denotes the MGF of the random variable $Z_\ell$ defined as
%$Z_\ell = |\alpha_{2,\ell}\exp(\jmath \phi_{2,\ell})- \alpha_{1,\ell}\exp(\jmath \phi_{1,\ell})|^2$,
$Z_\ell = |z_{2,\ell}- z_{1,\ell}|^2$ with $z_{i,\ell} = \alpha_{i, \ell}\exp(\jmath \Phi_{2,\ell})$ being random vectors having
arbitrarily distributed magnitudes $\alpha_{i, \ell}$ and phases $\Phi_{i,\ell}$, $\forall i \in \{1, 2\}$. %\footnote{Throughout this analysis and unless otherwise stated, indices $i$ and $\ell$ take the values $i = 1,2$ and $\ell = 1, \ldots, N_r$}.
In general, closed form expressions for $\mathcal{I}(A,L)$ are very difficult to be obtained and numerical integration is used instead (see for example \cite{J:RenzoNakagami}, \cite{J:RenzoRice} and \cite{J:Peppas_13}). In \cite{J:RenzoAsym}, by exploiting asymptotic analysis, closed-form approximations for $\mathcal{I}(A,L)$ are provided for high values of $A$, assuming that $\alpha_{i, \ell}$ are Nakagami-$m$ distributed random variables and $\Phi_{i,\ell}$ uniformly distributed in $[0, 2\pi]$.

In the following analysis, a generic solution of \eqref{Eq:EqIint} for high values of $A$ will be deduced, assuming that $\alpha_{i, \ell}$ are arbitrarily distributed random variables and $\Phi_{i,\ell}$ are uniformly distributed in $[0, 2\pi]$.
In order to obtain such an expression, \cite[Proposition 3]{J:WangGiannakis} is employed to approximate $\mathcal{M}_{Z_\ell}(s)$ for $s\rightarrow \infty$ as\footnote{The notation $f(x)=o[g(x)]$ as $x\rightarrow x_0$ stands for $\lim_{x\rightarrow x_0} \frac{f(x)}{g(x)} = 0$.}
\begin{align}\label{Eq:MGFTaylor}
|\mathcal{M}_{Z_{\ell}}(s)| = c_\ell|s|^{-d_\ell}+o(|s|^{-d_\ell}), \, s\rightarrow\infty
\end{align}
The MGF of $Z_{\ell}$ is given by \cite[Eq. (9)]{J:Peppas_13} as
\begin{equation}\label{Eq:MGF}
\mathcal{M}_{Z_{\ell}}(s) = \frac{1}{2s}
\int_0^{\infty}R \mathrm{e}^{-\frac{R^2}{4s}}\left[\prod_{i=1}^2\mathcal{H}_{0,R}\left\{\frac{f_{\alpha_{i,\ell}}(r)}{r}\right\}\right]\mathrm{d}R
\end{equation}
where $f_{a_{i,\ell}}(r)$ is the probability density function of $\alpha_{i,\ell}$ and $\mathcal{H}_{0,R}\{ \cdot\}$ denotes the zeroth order Hankel transform \cite[Eq. (9.11)]{B:Poularikas}.
Since $\mathrm{e}^{-\frac{R^2}{4s}} = o(\frac{1}{s})$ as $s\rightarrow\infty$, the approximation $\mathrm{e}^{-\frac{R^2}{4s}}/2s\overset{s\rightarrow\infty}{\approx} 1/2s$ can be employed in \eqref{Eq:MGF} to yield
\begin{equation}\label{Eq:MGF2}
\mathcal{M}_{Z_{\ell}}(s) \overset{s\rightarrow\infty}{\approx} \frac{1}{2s}
\int_0^{\infty}R \left[\prod_{i=1}^2\mathcal{H}_{0,R}\left\{\frac{f_{\alpha_{{i},\ell}}(r)}{r}\right\}\right]\mathrm{d}R
\end{equation}
Changing the information variable $R^2$ to $y$ and comparing \eqref{Eq:MGF2} with \eqref{Eq:MGFTaylor}, it is readily deduced that $d_\ell = 1$ and
\begin{equation}\label{Eq:codinggain}
c_\ell = \frac{1}{4}\int_{0}^{\infty}
\prod_{i=1}^2\mathcal{H}_{0,\sqrt{y}}\left\{\frac{f_{\alpha_{i,\ell}}(r)}{r}\right\}\mathrm{d}y
\end{equation}
Finally, by substituting \eqref{Eq:MGFTaylor} and \eqref{Eq:codinggain} into \eqref{Eq:EqIint}, it is deduced that for high values of $A$, $\mathcal{I}(A,L)$ can be approximated by
\begin{equation}\label{Eq:PEPasym}
\mathcal{I}(A,L) \overset{A \gg 1}{\approx} \frac{2^{L - 1}\Gamma\left(L + \frac{1}{2}\right)}{\sqrt{\pi}\Gamma\left(L + 1\right)} \left[\prod_{\ell=1}^{L} c_\ell\right] A^{-L}
\end{equation}
%$\int_0^{\infty}R \left[\prod_{i=1}^2\mathcal{H}_{0,R}\left\{\frac{f_{a_{t_{i},\ell}}(r)}{r}\right\}\right]\mathrm{d}R$.
%Changing the information variable $R^2$ to $y$, $c_\ell$ can be readily expressed as in \eqref{Eq:codinggain}.
%Then, by employing \eqref{Eq:MGFprod}, $|\mathcal{M}_{Z^2}(s)|$ can be written as
%\begin{equation}\label{Eq:MGFallTaylor}
%|\mathcal{M}_{Z^2}(s)| = \left[\prod_{\ell=1}^{N_r}c_\ell\right]|s|^{-N_r}+o(|s|^{-N_r}), \, s\rightarrow\infty,
%\end{equation}
%wherefrom it is deduced that the diversity and coding gains achieved by the considered system are $N_r$ and $\prod_{\ell=1}^{N_r}c_\ell$, respectively.
%in \eqref{Eq:PEPasym} which concludes the proof.
%Observe that the diversity gain depends only on the number of the receive antennas and is independent of the fading severity.
%This finding is in agreement with relevant findings reported in \cite{J:RenzoRice} and \cite{J:RenzoAsym}.
where $\Gamma(\cdot)$ is the gamma function \cite[Eq. (8.310/1)]{B:Gradshteyn_00}.

\newcounter{mytempeqncnt}
%_{e_\ell [r]}%
\begin{figure*}[!t]
\normalsize \setcounter{mytempeqncnt}{\value{equation}}
\setcounter{equation}{8}
\begin{equation}\label{Eq:CodinggainEGK}
\begin{split}
& c_\ell = \frac{b_{s,1,\ell}b_{1,\ell}}{\Gamma(m_{1,\ell})\Gamma(m_{2,\ell})\Gamma(m_{s,1,\ell})\Gamma(m_{s,2,\ell})\Omega_{1,\ell}} H\substack{ 2 , 2 \\ 2 , 2} \left[ \frac{\Omega_{2,\ell}b_{1,\ell}b_{s,1,\ell}}{\Omega_{1,\ell}b_{2,\ell}b_{s, 2,\ell}}
\left| \substack{ \left(1-m_{2,\ell},\frac{2}{\beta_{2,\ell}}\right),\left(1-m_{s,2,\ell},\frac{2}{\beta_{s,2,\ell}}\right) \\ \null\\
%\left(m_{i,\ell}, \frac{2}{\beta_{i,\ell}}\right), \left(m_{s, i,\ell}, \frac{2}{\beta_{s, i,\ell}}\right)
\left(m_{1,\ell}-\frac{2}{\beta_{1,\ell}},\frac{2}{\beta_{1,\ell}}\right),
\left(m_{s,1,\ell}-\frac{2}{\beta_{s,1,\ell}},\frac{2}{\beta_{s,1,\ell}}\right)
 }\right.\right]
\end{split}
\end{equation}
\hrulefill \vspace*{1pt} \setcounter{equation}{\value{mytempeqncnt}}
\end{figure*}
\pubidadjcol
\begin{figure*}[!t]
\normalsize \setcounter{mytempeqncnt}{\value{equation}}
\setcounter{equation}{10}
\begin{equation}\label{Eq:CodinggainKG}
\begin{split}
& c_\ell = \frac{\Gamma(-1+m_{2,\ell}+m_{s,1,\ell})\Gamma(-1+m_{s,2,\ell}+m_{s,1,\ell})\Gamma(-1+m_{2,\ell}+m_{1,\ell})\Gamma(-1+m_{s,2,\ell}+m_{1,\ell})
}
{\Gamma\left(-2+\sum_{i=1}^2\left[m_{i,\ell}+m_{s,i,\ell}\right]\right)\left[\prod_{i=1}^2\Gamma(m_{i,\ell})\Gamma(m_{s,i,\ell})\right]
\left(\frac{m_{s,1,\ell}m_{1,\ell}}{\Omega_{1,\ell}}\right)^{m_{2,\ell}-1}
}  \\
& \times \left(\frac{m_{2,\ell}m_{s,2,\ell}}{\Omega_{2,\ell}}\right)^{m_{2,\ell}}
\null_2F_1\left(-1+\sum_{i=1}^2m_{i,\ell}, -1+m_{2,\ell}+m_{s,1,\ell};-2+\sum_{i=1}^2\left[m_{i,\ell}+m_{s,i,\ell}\right],1-\frac{\Omega_{1,\ell}m_{2,\ell}m_{s,2,\ell}}{\Omega_{2,\ell}m_{1,\ell}m_{s, 1,\ell}}\right)
% H\substack{ 2 , 2 \\ 2 , 2} \left[ \frac{\Omega_{2,\ell}b_{1,\ell}b_{s,1,\ell}}{\Omega_{1,\ell}b_{2,\ell}b_{s, 2,\ell}}
%\left| \substack{ \left(1-m_{2,\ell},\frac{2}{\beta_{2,\ell}}\right),\left(1-m_{s,2,\ell},\frac{2}{\beta_{s,2,\ell}}\right) \\ \null\\
%%\left(m_{i,\ell}, \frac{2}{\beta_{i,\ell}}\right), \left(m_{s, i,\ell}, \frac{2}{\beta_{s, i,\ell}}\right)
%\left(m_{1,\ell}-\frac{2}{\beta_{1,\ell}},\frac{2}{\beta_{1,\ell}}\right),
%\left(m_{s,1,\ell}-\frac{2}{\beta_{s,1,\ell}},\frac{2}{\beta_{s,1,\ell}}\right)
% }\right.\right]
\end{split}
\end{equation}
\hrulefill \vspace*{1pt} \setcounter{equation}{\value{mytempeqncnt}}
\end{figure*}
\pubidadjcol

It is noted that $c_\ell$ and henceforth $\mathcal{I}(A,L)$
can be easily obtained in closed-form by employing Mellin transform techniques, provided that a closed-form expression for $\mathcal{H}_{0,\sqrt{y}}\left\{\frac{f_{\alpha_{i,\ell}}(r)}{r}\right\}$ is readily available. In what follows, a closed form expression for $c_\ell$ will be deduced assuming that $\alpha_{i, \ell}$ follow the Extended Generalized-$\mathcal{K}$ (EGK) distribution. The motivation behind the choice of this specific model is that the EGK distribution exhibits good tail properties and encompasses most of the well-known fading distributions
either as special or as limiting cases \cite[Table I]{C:Yilmaz}. %Its wide versatility renders it suitable for modeling fading in
%wireless millimeter wave (60 GHz and above) and free-space
%optical channels.
Simplified expressions for the special cases of Generalized-$\mathcal{K}$ and the Nakagami-$m$ distributions are also deduced.
\subsection{The Extended Generalized-$\mathcal{K}$ case}
Under EGK fading, the zeroth order Hankel transform of $f_{\alpha_{i,\ell}}(r)/r$ is determined in closed-form as \cite[Eq. (11)]{J:Peppas_13}
 \begin{equation}\label{Eq:EGKHankel}
\begin{split}
\mathcal{H}_{0,R}\left\{\frac{f_{\alpha_{i,\ell}}(r)}{r}\right\} &= \frac{H\substack{ 2 , 1 \\ 2 , 2} \left[ \frac{4b_{s, i,\ell}\,{b}_{i,\ell}}{R^2\Omega_{i,\ell}}
\left| \substack{ (1,1),\ (1,1) \\ \null\\
%\left(m_{i,\ell}, \frac{2}{\beta_{i,\ell}}\right), \left(m_{s, i,\ell}, \frac{2}{\beta_{s, i,\ell}}\right)
\Xi_\ell }\right.\right]}{\Gamma(m_{i,\ell})\Gamma(m_{s,i,\ell})}
\end{split}
\end{equation}
where $H \,\substack{ m , n \\ p , q}[\cdot]$ is the Fox's H-function \cite[Eq. (8.3.1)]{B:Prudnikov3}\footnote{Note that efficient algorithms for the numerical evaluation of the H-function are available in \cite[Table 2]{J:Peppas2012} and \cite[Appendix A]{C:Yilmaz4}.
.}, $\Xi_\ell \triangleq \left\{\left(m_{i,\ell}, \frac{2}{\beta_{i,\ell}}\right), \left(m_{s, i,\ell}, \frac{2}{\beta_{s, i,\ell}}\right)\right\}$.
In \eqref{Eq:EGKHankel}, $m_{i,\ell}$ $(0.5 < m_{i,\ell} < \infty)$ and $\beta_{i,\ell}$ $(0 < \beta_{i,\ell} <\infty)$ represent the fading
severity and the fading shaping factor, respectively, $m_{s,i,\ell}$ $(0.5 < m_{s,i,\ell} < \infty)$ and $\beta_{s,i,\ell}$ $(0<\beta_{s,i,\ell} <\infty)$ represent the shadowing severity and the shadowing shaping factor, respectively, and  $\Omega_{i,\ell}=\mathbb{E}\langle a_{i,\ell}^2 \rangle$ with $\mathbb{E}\langle \cdot \rangle$ denoting expectation. Moreover, ${b}_{i,\ell} = {\Gamma\left(m_{i,\ell}+\frac{2}{\beta_{i,\ell}}\right)}/{\Gamma(m_{i,\ell})}$ and $b_{s,i,\ell} = {\Gamma\left(m_{s,i,\ell}+\frac{2}{\beta_{s,i,\ell}}\right)}/{\Gamma(m_{s,i,\ell})}$.
%Under EGK fading conditions, $c_\ell$ can be expressed in closed form as \eqref{Eq:CodinggainEGK}, on the top of the next page.
By substituting \eqref{Eq:EGKHankel} into \eqref{Eq:codinggain} and employing %the well-known result stating that the Mellin transform of the product of two H-functions is also an H-function
\cite[Eq. (2.25.1.1)]{B:Prudnikov3} along with \cite[Eq. (8.3.2.7)]{B:Prudnikov3}, $c_\ell$ can be evaluated from
\begin{equation}\label{Eq:CodingGainInterm}
c_\ell = A_{\ell}H \substack{ 3 , 3 \\ 4 , 4} \left[ x_{\ell}
\left| \substack{ (\Lambda_1, \lambda_1), (\Lambda_2, \lambda_2), (0,1), (0,1)\\ \null\\
%\left(m_{i,\ell}, \frac{2}{\beta_{i,\ell}}\right), \left(m_{s, i,\ell}, \frac{2}{\beta_{s, i,\ell}}\right)
(0,1), (M_1, \mu_1), (M_2, \mu_2), (0,1)
 }\right.\right]
\end{equation}
where $A_{\ell} = \frac{b_{s,1,\ell}b_{1,\ell}}{\Gamma(m_{1,\ell})\Gamma(m_{2,\ell})\Gamma(m_{s,1,\ell})\Gamma(m_{s,2,\ell})\Omega_{1,\ell}}$,
$x_{\ell} = \frac{\Omega_{2,\ell}b_{1,\ell}b_{s,1,\ell}}{\Omega_{1,\ell}b_{2,\ell}b_{s, 2,\ell}}$,
$\Lambda_1 = 1-m_{2,\ell}$, $\lambda_1 ={2}/{\beta_{2,\ell}}$, $\Lambda_2 = 1-m_{s,2,\ell}$, $\lambda_2 ={2}/{\beta_{s,2,\ell}}$,
$M_1 = m_{1,\ell}-{2}/{\beta_{1,\ell}}$,  $\mu_1 = {2}/{\beta_{1,\ell}}$, $M_2 = m_{s,1,\ell}-{2}/{\beta_{s,1,\ell}}$
and $\mu_2 = {2}/{\beta_{s,1,\ell}}$.
The result in \eqref{Eq:CodingGainInterm} can be reduced further by employing \cite[Eq. (8.3.2.6)]{B:Prudnikov3}
yielding \eqref{Eq:CodinggainEGK}, on the top of the next page.
%if the H-function is
%expressed as a Mellin-Barnes contour integral \cite[Eq. (8.3.2.21)]{B:Prudnikov3} and simplifying
%the resulting product of gamma functions yielding
%\begin{align}\label{Eq:CodingGainInterm2}
% c_\ell = \frac{\mathcal{A}_\ell}{2\pi\jmath}\int_{\mathcal{C}}\prod_{i=1}^2\left[\Gamma\left(M_i+ \mu_iu\right)
%\Gamma\left(1-\Lambda_i- \lambda_iu\right)\right]
%{x_\ell}^{-u}\mathrm{d}u
%\end{align}
%where $\jmath = \sqrt{-1}$ and $\mathcal{C}$ is the Mellin-Barnes contour. The last integral is recognized as the definition of the H-function in
%\eqref{Eq:CodinggainEGK} and, therefore, $c_\ell$ can be expressed in closed form as
%\begin{equation}
%\begin{split}
%& \Lambda_\ell = \left\{,\left(1-m_{s,2,\ell},\frac{2}{\beta_{s,2,\ell}}\right)\right. \\
%& \left.,(0,1), (0,1)\right\} \\
%& M_\ell =
%\left(m_{1,\ell}-\frac{2}{\beta_{1,\ell}},\frac{2}{\beta_{1,\ell}}\right),
%\left(m_{s,1,\ell}-\frac{2}{\beta_{s,1,\ell}},\frac{2}{\beta_{s,1,\ell}}\right)
%\end{split}
%\end{equation}
%
%\end{IEEEproof}
 %Substituting \eqref{Eq:EGKHankel} into \eqref{Eq:ABEPParseval2}, the ABEP of SSK for MISO systems is readily evaluated
  \subsection{The Generalized-$\mathcal{K}$ case}
Under Generalized-$\mathcal{K}$ fading conditions, an expression for $c_\ell$
is readily obtained from \eqref{Eq:CodinggainEGK} setting the fading shaping factor $\beta_{i,\ell} \rightarrow 2$ and the shadowing shaping factor $\beta_{s, i,\ell} \rightarrow 2$.
Employing \cite[Eq. (8.3.2.21)]{B:Prudnikov3}, \eqref{Eq:CodinggainEGK} yields
\setcounter{equation}{9}
 \begin{equation}\label{Eq:KGcl}
\begin{split}
& c_\ell = \mathcal{B}_{\ell} G\substack{ 2 , 2 \\ 2 , 2} \left[ \frac{\Omega_{2,\ell}m_{1,\ell}m_{s,1,\ell}}{\Omega_{1,\ell}m_{2,\ell}m_{s, 2,\ell}}
\left| \substack{ 1-m_{2,\ell}, 1-m_{s,2,\ell} \\ \null\\
%\left(m_{i,\ell}, \frac{2}{\beta_{i,\ell}}\right), \left(m_{s, i,\ell}, \frac{2}{\beta_{s, i,\ell}}\right)
m_{1,\ell}-1, m_{s,1,\ell}-1 }\right.\right]
\end{split}
\end{equation}
where $G \,\substack{ m , n \\ p , q}[\cdot]$ is the Meijer's G-function \cite[Eq. (9.301)]{B:Gradshteyn_00} and $\mathcal{B}_{\ell} = \frac{m_{s,1,\ell}m_{1,\ell}}{\Gamma(m_{1,\ell})\Gamma(m_{2,\ell})\Gamma(m_{s,1,\ell})\Gamma(m_{s,2,\ell})\Omega_{1,\ell}}$. Finally, employing the identity \cite[Eq. (07.34.03.0871.01)]{Wolfram},
\eqref{Eq:KGcl} can be further expressed in terms of the Gauss hypergeometric function $\null_pF_q(\cdot)$ \cite[Eq. (9.14.1)]{B:Gradshteyn_00} as \eqref{Eq:CodinggainKG}, on the top of the next page.
\subsection{The Nakagami-$m$ case}
For the special case of Nakagami-$m$ fading, an expression for $c_\ell$ is readily obtained from \eqref{Eq:KGcl} setting the shadowing severity factor $m_{s,i,\ell} \rightarrow \infty$. Specifically, it can be shown that $c_\ell$ is reduced to a known result.
Letting $m_{s, i,\ell} \rightarrow \infty$ in \eqref{Eq:KGcl} and employing the definition of the Meijer's G-function \cite[Eq. (8.2.1.1)]{B:Prudnikov3},
$c_\ell$ can be written as
\setcounter{equation}{11}
\begin{align}\label{Eq:cellNakagami1}
%\begin{split}
& c_\ell  = \frac{m_{1,\ell}(2\pi\jmath \Omega_{1,\ell})^{-1}}{\Gamma(m_{1,\ell})\Gamma(m_{2,\ell})}
\int_{\mathcal{C}}\left(\frac{\Omega_{2,\ell}m_{1,\ell}}{\Omega_{1,\ell}m_{2,\ell}}\right)^{-u}\Gamma(m_{2,\ell}-u) \nonumber \\
& \times \Gamma(m_{1,\ell}-1+u)\left[\lim_{m_{s, 1,\ell}\to\infty}\frac{m_{s,1,\ell}^{-u+1}\Gamma(m_{s,1,\ell}-1+u)}{\Gamma(m_{s,1,\ell})}\right] \nonumber \\
& \times \left[\lim_{m_{s, 2,\ell}\to\infty}\frac{m_{s,2,\ell}^{u}\Gamma(m_{s,2,\ell}-u)}{\Gamma(m_{s,2,\ell})}\right]\mathrm{d}u
%\end{split}
\end{align}
where $\mathcal{C}$ is the Mellin-Barnes contour. Employing the identity $\lim_{x \to \infty}\frac{x^{-u}\Gamma(x+u)}{\Gamma(x)} = 1$ \cite[Eq. (8.328)]{B:Gradshteyn_00} along with \cite[Eq. (8.2.1.1)]{B:Prudnikov3},
\eqref{Eq:cellNakagami1} is written as
\begin{equation}\label{Eq:cellNakagami2}
\begin{split}
& c_\ell  = \frac{m_{1,\ell}}{\Gamma(m_{1,\ell})\Gamma(m_{2,\ell})\Omega_{1,\ell}}
G\substack{ 1 , 1 \\ 1 , 1} \left[ \frac{\Omega_{2,\ell}m_{1,\ell}}{\Omega_{1,\ell}m_{2,\ell}}
\left| \substack{  1-m_{2,\ell}\\ \null\\
m_{1,\ell}-1}\right.\right]
\end{split}
\end{equation}
Finally, using the identity $G\substack{ 1 , 1 \\ 1 , 1} \left[ x
\left| \substack{ a \\ \null\\
b}\right.\right] = \Gamma(1-a+b)x^b(x+1)^{a-b-1}$ \cite[07.34.03.0271.01]{Wolfram},
$c_\ell$ is given from
\begin{equation}\label{Eq:CodinggainNakagami}
\begin{split}
c_\ell &= \left[\prod_{i=1}^2\frac{1}{\Gamma(m_{i,\ell})}\left(\frac{m_{i,\ell}}{\Omega_{i,\ell}}\right)^{m_{i,\ell}}\right] \Gamma\left(-1+\sum_{i=1}^2m_{i,\ell}\right)\\
& \times \left(\sum_{i=1}^2\frac{m_{i,\ell}}{\Omega_{i,\ell}}\right)^{1-\sum_{i=1}^2m_{i,\ell}}
\end{split}
\end{equation}
which is identical to \cite[Eq. (4)]{J:RenzoAsym}.
\section{Application to the Performance analysis of Spatial Modulation}\label{Sec:Performance}
In this section, the results reported in Section~\ref{Sec:Tools} are applied to assess the asymptotic performance of SM systems.
\subsection{System Model}
A $N_t \times N_r$ MIMO system employing SM is considered, equipped with $N_t$ transmit and $N_r$ receive antennas, which can send digital information via $M$ complex symbols, $\chi_j = |\chi_j|e^{\jmath \theta_j}$,
$j = 1,\ldots, M$.
In the following and without loss of generality, two test cases are considered: \emph{i}) A pure SSK system operating under
independent and identically distributed (i.i.d) fading (\emph{Case I}); and \emph{ii}) A SM system operating under
i.i.d fading with constant-modulus modulation i.e. $|\chi_j| = \kappa_0$, $\forall j = 1,\ldots,M$ (\emph{Case II}).
\subsubsection{Case I}
Under the assumption of i.i.d fading, a tight upper bound for the ABEP of SSK can be obtained from \cite[Eq. (35)]{J:RenzoNakagami}, \cite{J:RenzoAsym}, as
\begin{equation}\label{Eq:ABEPMxN}
\overline{P} \leq \frac{N_t}{2}{\rm PEP_{SSK}}(t_1\rightarrow t_2)
\end{equation}
where ${\rm PEP_{SSK}}(t_1\rightarrow t_2)$ denotes the pairwise error probability related to the pair of transmit antennas $t_1$ and $t_2$,
$t_1, t_2 = 1, 2, \ldots, N_t$, and it is the
same for any pair $(t_1, t_2)$.
%The PEP can be evaluated as \cite[Eq. (11)]{J:RenzoRice}
%\begin{equation}\label{Eq:ABEP2xNrPDF}
%{\rm PEP}(t_1\rightarrow t_2) = \int_0^{\infty}Q\left(\sqrt{\overline{\gamma}}z\right)f_{Z}(z)dz
%\end{equation}
The ${\rm PEP_{SSK}}(t_1\rightarrow t_2)$ can be evaluated as \cite[Eq. (1)]{J:RenzoAsym}
\begin{equation}\label{Eq:ABEP2xNrMGF}
{\rm PEP_{SSK}}(t_1\rightarrow t_2) = \frac{1}{\pi}\int_0^{\pi/2}\prod_{\ell = 1}^{N_r}\left[\mathcal{M}_{\mathcal{Z}_\ell}\left(\frac{\overline{\gamma}}{2\sin^2\theta}\right)\right]\mathrm{d}\theta
\end{equation}
where $\mathcal{Z}_\ell = |a_{t_{2},\ell}\exp(\jmath \phi_{t_{2},\ell})- a_{t_{1},\ell}\exp(\jmath \phi_{t_{1},\ell})|^2$,
with $a_{t_i,\ell}$ and $\phi_{t_i,\ell}$ being the envelopes and phases of the link
defined by the $t_i$-th transmit antenna and the $\ell$-th receive antenna.
Moreover, $\overline{\gamma} = E_s/4N_0$ is the SNR where $E_s$ is the symbol energy and $N_0$ is the single-sided power spectral density of the additive white gaussian noise. For high values of $\overline{\gamma}$, it can be observed that ${\rm PEP_{SSK}}(t_1\rightarrow t_2)$ can be readily evaluated employing \eqref{Eq:PEPasym} as
${\rm PEP_{SSK}}(t_1\rightarrow t_2) \overset{\overline{\gamma} \gg 1}{\approx} \mathcal{I}\left(\overline{\gamma}, N_r\right)$.
Finally, from \eqref{Eq:PEPasym}, it is evident that the diversity gain depends only on the number of the receive antennas and is independent of the fading severity. This finding is in agreement with relevant findings reported in \cite{J:RenzoRice} and \cite{J:RenzoAsym}. The resulting coding gain can be obtained in closed-form from \cite[Eq. (1)]{J:WangGiannakis}.

\subsubsection{Case II}
The ABEP of SM can be tightly upper bounded as \cite[Eq. (6)]{J:SMRenzoHaasTVT}
\begin{align}\label{Eq:ABEPSMMxN}
\overline{P} &\leq \text{ABEP}_\text{signal}+\text{ABEP}_\text{spatial}+\text{ABEP}_\text{joint}
\end{align}
where $\text{ABEP}_\text{signal}$, $\text{ABEP}_\text{spatial}$ and $\text{ABEP}_\text{joint}$ show how the error performance of SM
is affected by the signal constellation diagram, the spatial constellation diagram and the interaction of
both signal and space constellation diagrams, respectively. Under generalized fading, the term $\text{ABEP}_\text{signal}$ when either $M$-ary phase shift keying ($M$-PSK) or $M$-ary quadrature amplitude modulation ($M$-QAM) are employed, can be readily evaluated using
\cite[Eqs. (7), (8)]{J:SMRenzoHaasTVT} and \cite[Table I]{J:SMRenzoHaasTVT}. High-SNR asymptotically tight expressions for
$\text{ABEP}_\text{signal}$ can also be obtained using \cite{J:WangGiannakis}.
Assuming constant modulus modulation $\text{ABEP}_{\text{spatial}}$ and $\text{ABEP}_{\text{joint}}$ can be obtained from \cite[Eq. (10)]{J:SMRenzoHaasTVT} and \cite[Eq. (11)]{J:SMRenzoHaasTVT}, respectively, as
\begin{subequations}
\begin{equation}
\text{ABEP}_{\text{spatial}} = \frac{N_t\log_2(N_t)}{2\log_2(N_tM)}{\rm PEP_{SM}}(t_1\rightarrow t_2)
\end{equation}
\begin{align}
%\begin{split}
\text{ABEP}_{\text{joint}} & = \left[
%\frac{M(N_t-1)\log_2(M)}{2\log_2(N_tM)}+\frac{N_t(M-1)\log_2(N_t)}{2\log_2(N_tM)}
\frac{M(N_t-1)\log_2(M)+N_t(M-1)\log_2(N_t)}{2\log_2(N_tM)}
\right] \nonumber \\
& \times{\rm PEP_{SM}}(t_1\rightarrow t_2)
%\end{split}
\end{align}
\end{subequations}
where ${\rm PEP_{SM}}(t_1\rightarrow t_2)$ can be readily obtained from \eqref{Eq:ABEP2xNrMGF} by replacing $\overline{\gamma}$ with $\kappa_0\overline{\gamma}$.
For high values of $\overline{\gamma}$, the framework presented in Section~\ref{Sec:Tools} can be readily employed to yield\footnote{When non-constant modulus modulation is assumed, the framework presented in Section~\ref{Sec:Tools} can be readily applied by setting in \eqref{Eq:EqIint} $\alpha_{i,\ell} = a_{t_i,\ell}|\chi_\ell|$ and
$\Phi_{i,\ell} = \phi_{t_i,\ell}+\theta_{\ell}$} ${\rm PEP_{SM}}(t_1\rightarrow t_2) \overset{\overline{\gamma} \gg 1}{\approx} \mathcal{I}\left(\kappa_0\overline{\gamma}, N_r\right)$.
The resulting diversity gain is $\min\{N_r, \text{Div}_{\text{signal}}\}$ where $\text{Div}_{\text{signal}}$ is the diversity gain of $\text{ABEP}_\text{signal}$ \cite{J:SMRenzoHaasTVT}.
\begin{figure}[!t]
\centering
\includegraphics[keepaspectratio,width=2.7in]{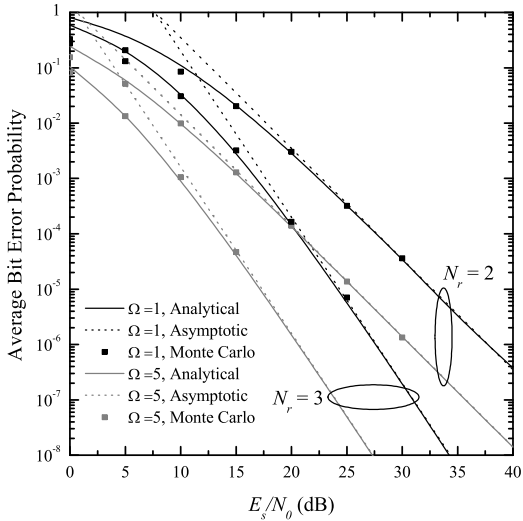}
\caption{ABEP of SSK for $8\times2$ and $8\times3$ MIMO EGK channels as a function of $E_s/N_0$.
Simulation Parameters: $m_{s,i,\ell} = 2$, $\beta_{s,i,\ell} = 1$,  $m_{i,\ell} = 1.5$, $\beta_{i,\ell} = 4$ and $\Omega_{i,\ell} \in \{1, 5\}$.
} \label{Fig:EGK}
\end{figure}
\begin{figure}[!t]
\centering
\includegraphics[keepaspectratio,width=2.7in]{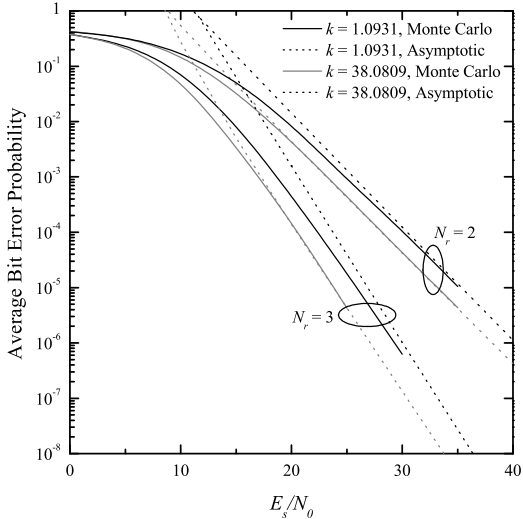}
\caption{ABEP of SM-QPSK for $8\times2$ and $8\times3$ MIMO Generalized-$\mathcal{K}$ channels as a function of $E_s/N_0$ for various values of $k$.
Simulation Parameters: $m_{i,\ell} = 1.5$, $\Omega_{i,\ell} = 1$.
} \label{Fig:KG}
\end{figure}
%\begin{figure}[!t]
%\centering
%\includegraphics[keepaspectratio,width=2.8in]{Figures/SSKSIMOKG}
%\caption{ABEP of SSK for $8\times1$ MISO Generalized-$\mathcal{K}$, channels, as a function of $E_s/N_0$
%for various values of $k$.
%Simulation Parameters: $\{m_{i}\}_{i=1}^{8} = i$, $\Omega_1 = 1$, $\{\Omega_{i}\}_{i=2}^{N_t} = 3(i-1)$
%} \label{Fig:SIMOKG}
%\end{figure}
\subsection{Numerical Results}
Numerical results accompanied by computer simulations are presented to
study the tightness of \eqref{Eq:PEPasym} under various fading conditions. In the following analysis, an $8\times N_r$ MIMO system is considered.
%For simplicity and without loss of generality, a fading scenario where all fading parameters are identical is considered, as it was done in \cite{J:RenzoAsym}. Since ${\rm PEP}(i \rightarrow t_j)$, $i,j=1,\ldots,N_t$ does not depend on the pair of the active antenna $i$ and the estimation $t_j$, the ABEP can be simplified as $\overline{P} \leq \left(N_t/2\right){\rm PEP}(1 \rightarrow 2)$, where ${\rm PEP}(1 \rightarrow 2)$ is given from \eqref{Eq:PEPasym}.
Fig.~\ref{Fig:EGK} depicts the ABEP of $8\times2$ and $8\times3$ MIMO SSK systems operating over EGK fading channels as a function of $E_s/N_0$,
assuming $m_{s,i,\ell} = 2$, $\beta_{s,i,\ell} = 1$,  $m_{i,\ell} = 1.5$, $\beta_{i,\ell} = 4$ and $\Omega_{i,\ell} \in \{1, 5\}$.
Fig.~\ref{Fig:EGK} includes upper bounds for the ABEP obtained by the numerical integration of
\eqref{Eq:ABEP2xNrMGF}, exact ABEP results obtained from Monte-Carlo simulation as well as asymptotic
ABEP results obtained employing the high SNR assumption in \eqref{Eq:PEPasym} and \eqref{Eq:CodinggainEGK}. As it is evident, the proposed analytical framework well predicts the diversity and coding gains of the considered system and yields tight results for high values of $E_s/N_0$. Moreover, it can be observed that $\Omega_{i,\ell}$ affects coding gain only and, as expected, coding gain improves as $\Omega_{i,\ell}$ increases.

For the same antenna configurations, Fig.~\ref{Fig:KG} depicts the ABEP of MIMO SM-QPSK $(M = 4)$, operating over generalized-$\mathcal{K}$ fading channels\footnote{Using \cite{J:WangGiannakis}, the diversity gain $\text{Div}_{\text{signal}}$ is deduced as $\min\{m_{s,i,\ell}, m_{i,\ell}\}$ } as a function of $E_s/N_0$, assuming $m_{i,\ell} = 1.5$, $\Omega_{i,\ell} = 1$ and $m_{s, i,\ell} = k$.
Different values of $k$ are considered to account for two shadowing scenarios, that is frequent heavy shadowing ($k = 1.0931$) and average shadowing
($k = 38.0809$) \cite{J:PeppasEGC}.
%Frequent Heavy
As for the tightness of \eqref{Eq:PEPasym}, similar conclusions to those reported in Fig.~\ref{Fig:EGK} are deduced. However, in the presence of heavy shadowing ($k = 1.0931$) and for $N_t = 3$ transmit antennas, the asymptotic behavior of the ABEP-SNR
curve shows up at high SNR values, i.e. for $E_s/N_0 > 30$dB. Furthermore, as it is expected, coding gain improves as $k$ increases, i.e. when the impact of shadowing becomes less severe.
%Finally, in Fig.~\ref{Fig:SIMOKG}, ABEP bounds for $8\times1$ MISO systems, obtained using \eqref{Eq:ABEPNx1},
%are plotted and compared with Monte Carlo simulations, assuming a generalized-$\mathcal{K}$ fading environment.
%and different shadowing scenarios. It is readily deduced that
%these bounds are accurate and asymptotically tight for all scenarios examined. These findings are in agreement with those reported in \cite{J:RenzoNakagami}.
\section{Conclusion} \label{Sec:Conclusion}
In this letter, an analytical framework for the
computation of the diversity and coding gains of SM systems
over generalized fading channels was presented. To the best of the authors'
knowledge, the derived Eqs. \eqref{Eq:codinggain}, \eqref{Eq:CodinggainEGK} and \eqref{Eq:CodinggainKG} are novel and
can be simplified to some particular cases already reported.
The newly derived simplified ABEP expressions require much less time for numerical evaluation
compared to the exact ones, which require numerical integration. It was shown that, under generalized fading,
the diversity gains of spatial and joint components of SM do not depend on the
fading severity.
%\bibliographystyle{IEEEtran}
%\bibliography{IEEEabrv,Peppas}

\begin{thebibliography}{10}
\providecommand{\url}[1]{#1}
\csname url@samestyle\endcsname
\providecommand{\newblock}{\relax}
\providecommand{\bibinfo}[2]{#2}
\providecommand{\BIBentrySTDinterwordspacing}{\spaceskip=0pt\relax}
\providecommand{\BIBentryALTinterwordstretchfactor}{4}
\providecommand{\BIBentryALTinterwordspacing}{\spaceskip=\fontdimen2\font plus
\BIBentryALTinterwordstretchfactor\fontdimen3\font minus
  \fontdimen4\font\relax}
\providecommand{\BIBforeignlanguage}[2]{{%
\expandafter\ifx\csname l@#1\endcsname\relax
\typeout{** WARNING: IEEEtran.bst: No hyphenation pattern has been}%
\typeout{** loaded for the language `#1'. Using the pattern for}%
\typeout{** the default language instead.}%
\else
\language=\csname l@#1\endcsname
\fi
#2}}
\providecommand{\BIBdecl}{\relax}
\BIBdecl

\bibitem{J:RenzoProceedingsIEEE}
M.~D. Renzo, H.~Haas, A.~Ghrayeb, S.~Sugiura, and L.~Hanzo, ``Spatial
  {M}odulation for {G}eneralized {MIMO}: {C}hallenges, {O}pportunities, and
  {I}mplementation,'' \emph{Proc. {IEEE}}, 2013.

\bibitem{J:SMRenzoSurvey}
M.~D. Renzo, H.~Haas, and P.~Grant, ``Spatial modulation for multiple-antenna
  wireless systems: a survey,'' \emph{IEEE Commun. Mag}, vol.~49, no.~12, pp.
  182--191, Dec. 2011.

\bibitem{J:RenzoNakagami}
M.~D. Renzo and H.~Haas, ``A general framework for performance analysis of
  {S}pace {S}hift {K}eying {(SSK)} modulation for {MISO} correlated
  {N}akagami-$m$ fading channels,'' \emph{{IEEE} Trans. Commun.}, vol.~59,
  no.~9, pp. 2590--2603, Sep. 2010.

\bibitem{J:RenzoRice}
------, ``Space {S}hift {K}eying ({SSK}–) {MIMO} over correlated {R}ician
  fading channels: {P}erformance analysis and a new method for
  transmit-diversity,'' \emph{{IEEE} Trans. Commun.}, vol.~59, no.~1, pp.
  116--129, Jan. 2011.

\bibitem{J:SMRenzoHaasTVT}
M.~D. Renzo, H.~Haas, and P.~Grant, ``Bit error probability of {SM}-{MIMO} over
  generalized fading channels,'' \emph{{IEEE} Trans. Veh. Technol.}, vol.~61,
  no.~3, pp. 1124--1144, Mar. 2012.

\bibitem{J:Peppas_13}
K.~Peppas, M.~Zamkotsian, F.~Lazarakis, and P.~Cottis, ``Unified error
  performance analysis of space shift keying modulation for {MISO} and {MIMO}
  systems under generalized fading,'' \emph{IEEE Wireless Commun. Letters},
  vol.~2, no.~6, pp. 663--666, 2013.

\bibitem{J:RenzoAsym}
M.~D. Renzo and H.~Haas, ``Bit error probability of space modulation over
  {N}akagami-$m$ fading: Asymptotic analysis,'' \emph{{IEEE} Commun. Lett.},
  vol.~15, no.~10, pp. 1026--1028, Oct. 2011.

\bibitem{J:WangGiannakis}
Z.~Wang and G.~Giannakis, ``A simple and general parametrization quantifying
  performance in fading channels,'' \emph{{IEEE} Trans. Commun.}, vol.~51,
  no.~8, pp. 1389--1398, Aug. 2003.

\bibitem{B:Poularikas}
A.~D. Poularikas, \emph{The Transforms and Applications Handbook},
  2nd~ed.\hskip 1em plus 0.5em minus 0.4em\relax CRC and IEEE Press, 2000.

\bibitem{B:Gradshteyn_00}
I.~Gradshteyn and I.~M. Ryzhik, \emph{Tables of Integrals, Series, and
  Products}, 6th~ed.\hskip 1em plus 0.5em minus 0.4em\relax New York: Academic
  Press, 2000.

\bibitem{C:Yilmaz}
F.~Yilmaz and M.-S. Alouini, ``A new simple model for composite fading
  channels: Second order statistics and channel capacity,'' in \emph{Proc.
  (ISWCS)}, York, Sep. 2010, pp. 676 -- 680.

\bibitem{B:Prudnikov3}
A.~P. Prudnikov, Y.~A. Brychkov, and O.~I. Marichev, \emph{Integrals and
  {S}eries {V}olume 3: {M}ore {S}pecial {F}unctions}, 1st~ed.\hskip 1em plus
  0.5em minus 0.4em\relax Taylor And Francis Ltd, 2003.

\bibitem{J:Peppas2012}
K.~Peppas, F.~Lazarakis, A.~Alexandridis, and K.~Dangakis, ``Simple, accurate
  formula for the average bit error probability of multiple-input
  multiple-output free-space optical links over negative exponential turbulence
  channels,'' \emph{Optics Letters}, vol.~37, pp. 3243--3245, Aug. 2012.

\bibitem{C:Yilmaz4}
F.~Yilmaz and M.-S. Alouini, ``Product of the powers of generalized
  {N}akagami-$m$ variates and performance of cascaded fading channels,'' in
  \emph{IEEE Global Telecommunications Conf.}, 2009, pp. 1 -- 8.

\bibitem{Wolfram}
T.~{W}olfram~function site, [{O}nline], available at
  \texttt{http://functions.wolfram.com}.

\bibitem{J:PeppasEGC}
K.~Peppas, ``Accurate closed-form approximations to generalized-{K} sum
  distributions and applications in the performance analysis of equal gain
  combining receivers,'' \emph{IET Commun.}, vol.~5, pp. 982--989, 2011.

\end{thebibliography}

\end{document}